# Thiol post-translational modifications modulate allosteric regulation of the OpcA-G6PDH complex through conformational gate control

**Running title: Thiol PTMs control the allostery of OpcA-G6PDH**

Hoshin Kim[1, #, *], Song Feng[2, #], Pavlo Bohutskyi[2,3], Xiaolu Li[2] , Daniel Mejia-Rodriguez[1], Tong Zhang[2], Wei-Jun Qian[2], and Margaret S. Cheung[4,5*]

[1]Physical Sciences Division, Physical and Computational Sciences Directorate, Pacific Northwest National Laboratory, Richland, Washington, USA

[2]Biological Sciences Division, Earth and Biological Sciences Directorate, Pacific Northwest National Laboratory, Richland, Washington, USA

[3]Department of Biological Systems Engineering, Washington State University, Pullman, Washington, USA

[4]Environmental Molecular Sciences Laboratory, Richland, Washington, USA

[5]University of Washington, Seattle, Washington, USA

[#] shares co-first authorship of this article.

[*] shares co-corresponding authorship of this article: hoshin.kim@pnnl.gov, margaret.cheung@pnnl.gov

# ABSTRACT

      Cyanobacteria require ultra-fast metabolic switching to maintain reducing power balance during environmental fluctuations. Glucose-6-phosphate dehydrogenase (G6PDH), catalyzing the rate-limiting step of the oxidative pentose phosphate pathway (OPPP), provides essential NADPH and metabolic intermediates for biosynthetic processes and redox homeostasis. In cyanobacteria, the unique redox-sensitive protein OpcA acts as a metabolic switch for G6PDH, enabling rapid adjustment of reducing power generation from glycogen catabolism and resulting in precise regulation of carbon flux between anabolic and catabolic pathways. While the redox-sensitive cysteine structures of OpcA are known to regulate G6PDH, the detailed mechanisms of how redox post-translational modifications (PTMs) influence OpcA's allosteric effects on G6PDH structures and function remain elusive. To investigate this mechanism, we utilized computational modeling combined with experimental redox proteomics using *Synechococcus elongatus* PCC 7942 as a model system. Redox proteomics captured modified cysteine residues under light/dark or circadian shifts. Computational simulation revealed that thiol PTMs near the OpcA-G6PDH interface are crucial to allosteric regulation of regions affecting the G6PDH activity, including a potential gate region for substrate ingress and product egress, as well as critical hydrogen bond networks within the active site. These PTMs promote rapid metabolic switching by enhancing G6PDH catalytic activity when OpcA is oxidized. This study provides evidence for novel molecular mechanisms that elucidate the importance of thiol PTMs of OpcA in modulating G6PDH structure and function in an allosteric manner, demonstrating how PTM-level regulation provides a critical control mechanism that enables cyanobacteria to rapidly adapt to environmental fluctuations through precise metabolic fine-tuning.

# INTRODUCTION

In cyanobacteria, the oxidative pentose phosphate pathway (OPPP) serves as a critical metabolic route for generating NADPH and biosynthetic precursors,(Diamond et al. 2017; Yang et al. 2002) particularly during periods when photosynthetic reducing power is unavailable (*e.g.,* dark conditions) or when cells initiate the restart of photosynthesis.(Makowka et al. 2020; Shinde et al. 2020) This pathway is essential for maintaining cellular redox homeostasis when photosynthesis cannot provide sufficient NADPH and biosynthetic intermediates for cellular processes and antioxidant systems. The first and rate-limiting enzyme of this pathway, glucose-6-phosphate dehydrogenase (G6PDH), is subject to intricate regulation to ensure adequate NADPH production while coordinating carbon flux between the OPPP and the Calvin-Benson cycle, thereby maintaining optimal reducing power balance and biosynthetic intermediate availability under varying environmental conditions.(Doello et al. 2025; Ito and Osanai 2020) A key regulatory mechanism involves the redox-sensitive oxidative pentose phosphate cycle protein (OpcA), which modulates G6PDH activity in response to the cellular redox state.(Doello et al. 2024; Née et al. 2013) During the daytime, when photosynthesis provides abundant reducing power and the Calvin-Benson cycle is active, OpcA forms a reduced state due to the action of the thioredoxin (Trx) system that obtains electrons from the ferredoxin pool. In this reduced form, OpcA inhibits G6PDH, minimizing the activity of the OPPP pathway to prevent unnecessary NADPH production and allowing only the minimal flux needed to sustain the OPPP shunt, which provides essential metabolic intermediates for biosynthetic processes while directing photosynthetically produced sugars into glycogen storage under stable light conditions. At night, however, cells rely on glycogen breakdown for energy, and the absence of photosynthetic electron transport leaves the Trx system in an oxidized state without reduced ferredoxin. Under these conditions, OpcA shifts to its oxidized form, which is critical for the rapid activation of cellular reducing power generation through the OPPP.(Doello et al. 2025)

Recent structural studies have shed light on the molecular basis of this regulation. Cryo-electron microscopy (cryo-EM) analyses revealed that oxidized OpcA binds to the G6PDH tetramer, inducing conformational changes that dramatically enhance enzymatic activity to meet the cell's immediate demand for reducing power.(Doello et al. 2024) Conversely, when photosynthetic reducing power becomes available, reduction of OpcA by the light-activated Trx system disrupts this interaction, leading to decreased G6PDH activity and effectively shutting down NADPH production via the OPPP. This redox-dependent modulation creates a sophisticated reducing power management system: under light conditions, photosynthetic electron transport provides abundant NADPH, allowing the Calvin-Benson cycle to operate with minimal flux through the OPPP shunt while directing the majority of fixed carbon toward glycogen storage for future energy and reducing power needs.(Johnson et al. 2025)

The redox sensitivity of OpcA is primarily mediated by intramolecular disulfide bridges involving conserved cysteine residues. The recent cryo-EM study of the OpcA-G6PDH complex from the cyanobacterial strain *Synechocystis sp.* PCC 6803 revealed that intramolecular disulfide bond formations in OpcA induce structural changes in the active site of G6PDH, enhancing its affinity for glucose-6-phosphate and increasing its enzymatic activity.(Doello et al. 2024) Mutagenesis and mass spectrometry analyses have identified Cys393 and Cys399 in OpcA from *Anabaena sp. PCC 7120* as critical residues for this redox regulation.(Mihara et al. 2018) Reduction of these disulfide bonds by thioredoxin leads to conformational alterations that prevent OpcA from effectively binding G6PDH, thereby rapidly attenuating its activity and shutting down NADPH production when reducing power is abundant from photosynthesis. These findings establish the molecular foundation for understanding how specific cysteine residues function as redox-sensitive switches in metabolic regulation. Interestingly, in heterocysts—specialized nitrogen-fixing cells in filamentous cyanobacteria—thioredoxin target proteins, including OpcA, remain more oxidized even under light conditions. This adaptation allows G6PDH to remain active in heterocysts during daylight, ensuring a continuous supply of reducing power for the highly energy-demanding process of nitrogen fixation.(Mihara et al. 2019) This specialized regulation demonstrates how cyanobacteria have fine-tuned the OpcA-G6PDH system to meet the distinct metabolic demands of different cell types within the same organism.

Despite these insights, experimentally capturing the transient and reversible nature of redox PTMs remains challenging. Traditional structural biology techniques, such as X-ray crystallography and nuclear magnetic resonance (NMR) spectroscopy, often fail to resolve the dynamic conformational states induced by PTMs, especially when modifications occur at low stoichiometry or are transient.(Chapman et al. 2023; Keedy et al. 2015) Mass spectrometry excels at identifying PTM sites but lacks the capacity to provide detailed structural information.(Azevedo et al. 2022) These limitations necessitate complementary approaches to fully understand the structural and functional ramifications of redox PTMs, particularly for understanding how these modifications enable rapid and targeted metabolic responses to environmental changes that are essential for cyanobacterial survival in fluctuating conditions.

Mass-spectrometry-based techniques have advanced to enable site-specific and stoichiometric analyses of cysteine modifications, such as *S*-glutathionylation, *S*-nitrosylation, and disulfide bond formation.(Li et al. 2021; Li et al. 2022) Innovative methodologies facilitate the enrichment and detection of reversibly oxidized cysteine residues, thereby illuminating the redox landscape within cells and tissues.(Guo et al. 2014; Li et al. 2023) These proteomic strategies are particularly valuable for uncovering redox-sensitive proteins and pathways that may be elusive to traditional biochemical assays(Duan et al. 2017; Fu et al. 2020; Zhang et al. 2021) and have proven especially powerful for characterizing dynamic redox changes in cyanobacteria during light/dark transitions that drive metabolic reprogramming. The computational workflow of **P**ost-**T**ranslational **M**odification on **P**rotein **S**tructures and their **I**mpacts on dynamics and functions (PTM-Psi) has been developed to leverage molecular dynamics (MD) simulations and advancements in force field development as a powerful tool to bridge this knowledge gap regarding how chemical modifications, such as PTMs, affect protein dynamics.(Mejia-Rodriguez et al. 2023) The PTM-Psi toolkit streamlines the parameterization of oxidized cysteine residues with quantum chemistry packages, enabling the investigation of their impact on protein structure and function. By providing atomistic insights into protein dynamics, MD simulations allow for the exploration of conformational landscapes inaccessible to static experimental methods. Furthermore, methods combining fluctuation relations with MD simulations have been developed to calculate redox potential changes in proteins, offering a computational approach to study redox regulation mechanisms.(Mejia-Rodriguez et al. 2023; Oliveira et al. 2024) Complementing MD simulations, redox proteomics has become an indispensable approach for the comprehensive identification and quantification of redox PTMs across the proteome.

      Here, we implement the complementary approach of redox proteomics and PTM-Psi to elucidate the conformational transitions associated with the redox state of OpcA and its binding affinity to G6PDH in the binary OpcA–G6PDH complex. Such studies provide valuable insights into the allosteric mechanisms governing the OPPP in cyanobacteria by revealing how rapid environmental responses are achieved through precise molecular switches operating on sub-second timescales to maintain cellular reducing power homeostasis. Moreover, integrating MD simulations with experimental data can facilitate the development of predictive models for redox regulation, with potential applications in metabolic engineering and synthetic biology. The dynamic interplay between redox PTMs and protein function, exemplified by the OpcA–G6PDH complex, highlights how computational approaches like MD simulations contribute to unraveling intricate regulatory mechanisms evolved in phototrophs for rapid response to environmental fluctuations to maintain metabolic efficiency. This understanding opens new avenues for developing predictive redox regulation models to enable rational metabolic interventions, with broad therapeutic and industrial applications.

## RESULTS

*Thiol PTMs within OpcA modulate protein structure and flexibility to enable metabolic switching*

The OpcA protein in *S. elongatus* comprises 445 amino acids (**Figure S1**) and contains four functionally important regions: (1) a peptidoglycan binding region (residues 55-106), (2) the G6PDH N-terminal domain (residues 131-247) (3) the G6PDH C-terminal domain (residues 255-432) and within this C-terminal domain, (4) the binding sites (residues 377-404) with G6PDH (**Figure 1A**). Regions (1)-(3) were annotated in UniProt (ID: Q54709)(Consortium 2024) and (4) was identified by a recent cryo-EM study.(Doello et al. 2024) Notably, all cysteine residues in the OpcA protein are strategically positioned within the putative G6PDH binding domain (**Figure 1A**), suggesting that redox-dependent oxidation of these cysteine residues directly modulates the formation and stability of the binary G6PDH-OpcA complex to control cellular reducing power generation through the OPPP.

We first performed all-atom MD simulations of the OpcA protein as a monomer with various cysteine PTMs to elucidate how these modifications might affect its structure and dynamics in the context of G6PDH regulation. In the case of OpcA with reduced cysteines (Reduced case), three flexible regions were observed, corresponding to amino acids ranging from 35-45, 110-125, and 160-174 (**Figure 1A**). Interestingly, our MD simulation results demonstrated that the region corresponding to the potential G6PDH binding site becomes structurally affected when three different types of PTM are applied to the cysteine sites, which was not observed in the reduced case (highlighted in red in **Figure 1B**). For example, this G6PDH binding site exhibited relatively large fluctuations when C398 was replaced with its glutathionylated form.

We predicted the structure of the OpcA-G6PDH complex using AlphaFold3(Abramson et al. 2024) to understand how OpcA interacts with the G6PDH complex, and to explore how possible cysteine PTMs can impact these protein-protein interactions (**Figure 2**). In this model, the complex is composed of four G6PDH subunits and one OpcA monomer, because OpcA monomers are known to interact with either a dimer of two identical G6PDH monomers or a tetramer of two dimers, depending on the conditions(Au et al. 2000; Kiani et al. 2007). Moreover, the only available 3D structure of this complex that can be compared with our model is from different cyanobacteria and consists of the tetrameric form of G6PDH complexed with an OpcA monomer.(Doello et al. 2024) The predicted monomeric structures of OpcA and G6PDH, as well as their complex form, showed high structural similarities when compared with the experimentally verified 3-D structures of OpcA(Doello et al. 2024) and G6PDH(Doello et al. 2024; Wei et al. 2022) (**Figure S2**). The predicted structures highlighted that the flexible regions influenced by PTMs, including regions I and III (highlighted in blue and purple in **Figure 1**, respectively) and the putative binding region (highlighted in red in **Figure 1**), are located near the interfacial region between G6PDH and OpcA. Specifically, the two flexible regions are positioned in the vicinity of the grooves of subunits A and C (**Figure 2A**), while the putative binding region forms interactions at the grooves of subunits A and B (**Figure 2B**). Additionally, the OpcA monomer makes close contact with three G6PDH subunits (A through C), while no direct contact with subunit D was observed (**Figure 2C**). Given that the amino acids within these flexible regions and putative binding sites are indeed involved in interactions with the G6PDH complex, we hypothesize that PTMs of the cysteine residues located within these two regions (residues 160-174 and 377-404) could affect the structure and function of the G6PDH complex.

*Redox proteomics data and computational analyses reveal the specific cysteine locations and types of PTMs that occur under physiological conditions*

To explore how light/dark perturbations and circadian rhythms affect the PTMs of these cysteines, we investigated all cysteine sites in OpcA using experimental redox proteomics data and calculated RBFE. The light-to-dark transition experiments were performed using both dense and dilute cultures adapted to continuous light conditions, ensuring that the observed redox changes are directly attributable to light availability without impact from circadian regulation. As illustrated in **Figure 3**, our redox proteomics data demonstrated that cysteines in flexible regions, including C162 and C174 (flexible region III, see **Figure 1A**) and C380, C386, and C398 (the putative G6PDH binding region, see **Figure 1A**), are sensitive to light-to-dark transitions and become increasingly oxidized under dark conditions when photosynthesis-driven reducing power becomes unavailable (**Figure 3A**). Importantly, OpcA protein abundance remained unchanged during the 2-hour light-to-dark transition in both dense and dilute cultures, indicating that PTM-based regulation operates independently of protein expression changes during short-term light perturbation. Cysteine residues C162–C174 and C380–C386 can form disulfide bonds (**Figure S1**) and, according to a recent cryo-EM study, these disulfide bonds allosterically influence the structure of the G6PDH complex and thus play a pivotal role in its functionality.(Doello et al. 2024) Notably, experimental data indicated that changes in oxidation of these cysteine residues were more pronounced when cells were grown in dilute cultures, where individual cells experienced higher levels of irradiation per cell due to reduced self-shading before transitioning to dark conditions. Furthermore, we discovered that glutathionylation at C398 is the most energetically favorable PTM compared to other potential thiol PTM sites during light/dark perturbations. This cysteine residue is not conserved in OpcA proteins from other species, highlighting its specificity to *S. elongatus* (**Figure S3**).

To examine the influence of circadian regulation beyond immediate light/dark responses, we compared the redox state of OpcA cysteines in diel cycle-adapted cells at two different circadian time points under identical light conditions. Importantly, in contrast to the short-term light/dark experiments where OpcA protein abundance remained unchanged, OpcA protein abundance showed significant changes between dusk and noon samples, revealing that abundance-based regulation operates on circadian timescales to modulate OpcA function. The redox state of cysteine residues showed distinct patterns when comparing dusk (8:00 p.m.) and noon (12:00 p.m.) samples, demonstrating clear differences between the light/dark transition experiments. C162 and C174 showed higher oxidation at dusk (8:00 p.m.) compared to noon (12:00 p.m.), while C380, C386, and C398 exhibited more complex patterns with insignificant changes for C380, substantial reduction for C386, and insignificant oxidation for C398, indicating that circadian and light/dark responses involve different cysteine modification patterns.

Overall, the computational analyses (**Figures 1, 2, and 3**), combined with the experimental redox proteomics data (**Figure 3**), highlight three key findings: first, two regions within OpcA (160-174 and 377-404) exhibit flexibility that is highly affected by thiol PTMs and are located at the grooves of the G6PDH subunits; second, five cysteine sites in these regions (C162, C174, C380, C386, and C398) tend to be oxidized under dark conditions; and third, the two disulfide bonds in these regions of OpcA, corresponding to C162-C174 and C380-C386, play a pivotal role in the activity of G6PDH in different cyanobacteria,(Doello et al. 2024) while glutathionylation at C398 represents one of the most energetically favorable PTMs. Given that these three modifications occur near the interfacial region between the OpcA and G6PDH in their complex, these PTMs likely significantly affect the structure and function of G6PDH. Based on our experimental redox data and computational analysis, as well as a recent cryo-EM study of other cyanobacteria,(Doello et al. 2024) we focused on the cysteine residues in the two interfacial regions of OpcA at the interface with G6PDH (C162, C174, C380, C386, and C398). We constructed and simulated two different OpcA–G6PDH complex systems: one with completely reduced cysteines without any PTMs or disulfide bonds, and another with two disulfide bonds (C162–C174 and C380–C386) and one glutathionylated PTM at C398. These simulations were designed to test our hypothesis regarding the impact of cysteine PTMs in OpcA on the structure and function of the G6PDH complex. Hereafter, we refer to these systems as the reduced and PTMed complexes, respectively.

*Thiol PTMs of OpcA in proximity to the G6PDH complex potentially regulate substrate access through allosteric gate modulation*

Our simulations demonstrated that stronger protein-protein interactions were observed between the OpcA monomer and G6PDH tetramer in the PTMed complex, exhibiting an increase of about 10 kcal/mol compared to the reduced complex (**Figure S4**). Furthermore, overall structural flexibility was reduced in both OpcA and G6PDH upon PTMs (**Figure S5**). This indicates that cysteine PTMs in the critical regions of OpcA enhance interactions with G6PDH and stabilize the entire complex structure through reduced fluctuations. Our simulation results also showed that PTMs significantly influence the structures and dynamics of regions near the active site of G6PDH (**Figure 4**). This region appears to function as a molecular gate, controlling the ingress of substrates or cofactors and the egress of reaction products. In the reduced complex, the gate adopts either a closed or wide-open conformation (**Figure 4A**), whereas PTMs modulate the gate conformations, resulting in more stable open conformations upon PTMs (**Figure 4B**). By analyzing the distance between the upper and lower sides of the gate in cryo-EM-based 3D structures of the G6PDH complex from different cyanobacteria, which range from 0.4 to 0.7 nm (**Figure 4C and D**), we categorized the G6PDH structures into three groups: "Open" conformations, where the majority of data points fall within this range; "Closed" conformations where the distances are less than 0.4 nm; and "Wide-open" conformations where the majority of distances exceed 0.7 nm (**Figure 4**). In the reduced complex, two of the four G6PDH subunits showed a "closed" gate, leading to reduced inner space near the catalytic cavity; one subunit showed alternating "open/closed" gate conformations with limited void volume near the active site; and one adopted a "wide-open" conformation with a large space in the cavity. However, when OpcA was PTMed, three out of four G6PDH subunits showed "open" conformations, and one exhibited an "open/closed" conformation. Subunit D, which in both the PTMed and reduced cases showed either "closed" or "open/closed" gate conformations, lacks close contacts with OpcA when assembled (**Figure 2C**), suggesting that interactions with OpcA indeed influence the complex's structure.

*Thiol PTMs of OpcA allosterically modulate gate conformations and hydrogen bond networks within the active site*

The upper and lower regions of the gate are composed of pairs of charged amino acids, ASP33 and ARG37, for the lower part, and GLU241 and ARG243 for the upper part (**Figure S1 and Figure 5**). Our simulations showed that the open and closed configurations arise from hydrogen bonding between ASP33 and ARG243. In the closed state, these two amino acids move closer and tend to form hydrogen bonds (**Figure 5B and E**), whereas in the open or wide-open states, they are positioned farther apart, resulting in minimal or no hydrogen bonding (**Figure 5C, D, and E**). Additionally, a hydrogen bond between GLU241 and ARG243 can also modulate the gate configurations. Unlike the consistent hydrogen bonding between ASP33 and ARG37, GLU241 and ARG243 bonds are formed in the closed conformation but are absent in the open or wide-open configurations (**Figure 5E**). Interestingly, when the gate is excessively wide-open, the local structure near the active site is disrupted, leading to the loss of crucial hydrogen bonds between active site residues, including ASP197 and HIS260 (**Figures 5D, 5E, and S1**). Similar gate mechanisms that directly affect enzyme activity and reaction rates have been reported across various biocatalytic systems.(Gora et al. 2013; Kokkonen et al. 2018) For example, recent combined experimental and computational studies have shown that enzyme activity and reaction rates are optimized when the ingress and egress gates remain in the "open" conformation. Conversely, enzyme activity and catalytic rates can be significantly reduced when the gate is "closed" or "wide-open".(Kim et al. 2016; Kim et al. 2014) This indicates that maintaining a stable open conformation ensures optimal enzyme activity, while activity can be significantly diminished in the closed state due to restricted substrate or product ingress and egress, or in the wide-open state due to unstable interactions between active site residues. Given that the residues involved in the gate configurations and the active site are conserved (**Figure S3**), the proposed gate mechanism could potentially play a similar role in G6PDH complexes across different organisms (*e.g.*, humans), regardless of whether they require OpcA.

As illustrated in **Figure S6**, the two complexes exhibited different allosteric interactions. In the reduced complex, only the gate residues in subunit A were allosterically affected by OpcA, and no long-range (~1.5 nm) contacts were observed between OpcA and subunits B, C, and D regarding gate configurations. In contrast, when OpcA was PTMed, long-range contacts were identified between OpcA and subunits A, B, and C, and the number of interactions with ARG243 increased. For subunit D, no allosteric effect by OpcA was evident in either case because there was no direct contact between the two proteins (**Figure 2C**).

## DISCUSSION

*Thiol PTMs regulate molecular assembly, energy production, and metabolic flux*

In cyanobacteria, G6PDH is not directly redox-sensitive, and its activity is modulated by thiol PTMs of the OpcA protein that binds to the G6PDH complex.(Doello et al. 2024) Experimental studies have demonstrated that the activity of G6PDH is drastically reduced in the absence of OpcA.(Hagen and Meeks 2001) Notably, the oxidized form of OpcA can lower the $K_m$ for G6P, highlighting the importance of the redox state of OpcA for the G6PDH enzyme function and cellular capacity to generate NADPH when reducing power is needed.(Mihara et al. 2018) More specifically, a recent cryo-EM study revealed two crucial intramolecular disulfide bonds of OpcA in the cyanobacterial strain *Synechocystis* sp., which play an essential role in the function of G6PDH proteins, acting as an allosteric activator.(Doello et al. 2024) Our simulation results, combined with redox proteomics, also demonstrate that these two disulfide bonds of OpcA influence the structural dynamics of the G6PDH-OpcA complex as well as the G6PDH enzyme activity. In addition, glutathionylated Cys398 also affects complex structure, dynamics, and function, which was not observed in previous studies (**Figures 3, 4, and 5**). Using the PTM-Psi tool we developed,(Mejia-Rodriguez et al. 2023) we discovered for the first time that the activity of the G6PDH-OpcA complex in the OPPP pathway is strongly linked to structural changes in the active site of G6PDH subunits induced by thiol PTMs (**Figures 4, 5, and S5**). When OpcA is reduced—where two disulfide bonds and glutathionylated cysteine are not present—the active site pocket in G6PDH exhibits either "closed" or "wide-open" configurations, which are not optimal for substrate binding or enzyme reactions. In contrast, upon oxidation, the active site pockets are predominantly maintained in optimal open configurations, with the exception of one G6PDH subunit that lacks direct interactions with OpcA. This study clearly indicates that PTMs at the OpcA-G6PDH interface are crucial for regulating the conformation and stability of active site regions in G6PDH, demonstrating how allosteric interactions from oxidized OpcA enable fine-tuning of enzymatic activity. Such redox-sensitive PTMs function as molecular switches that allow rapid response to environmental changes, which is particularly important for metabolically versatile organisms like cyanobacteria.

*Hierarchical regulation of cellular reducing power production and carbon flux through dual temporal mechanisms*

The primary biological function of the OpcA-G6PDH system is maintaining cellular reducing power balance through NADPH generation when photosynthetic electron transport is unavailable or insufficient. The formation and reduction of cysteine-based PTMs occur on the sub-second-to-minute time scale—several orders of magnitude faster than transcriptional or translational control.(Doello et al. 2024; Johnson et al. 2025; Li et al. 2023) Within this kinetic window, OpcA operates as a molecular "rheostat" that couples the cellular redox state to immediate NADPH production through the OPPP. When light intensity drops or reactive-oxygen species accumulate, thiol oxidants promote rapid intramolecular disulfide formation. The oxidized form of OpcA engages G6PDH with high affinity, boosting NADPH production to meet urgent cellular reducing power demands. The additional reducing power is immediately channeled to (1) antioxidant systems such as peroxiredoxins and glutathione reductase, and (2) anabolic reactions that rely on NADPH (*e.g.*, $CO_2$ fixation and glycogen synthesis). Conversely, re-reduction of the disulfide by thioredoxin, itself fueled by photosynthetic ferredoxin, makes OpcA-G6PDH inactive within seconds after light returns. This "on–off" cycle prevents futile NADPH overproduction when photosynthesis can adequately supply reducing power and redirects glucose-6-phosphate back into glycolysis and the Calvin–Benson-Bassham (CBB) cycle. Within this cycle, the molecular gate conformation, regulated by OpcA, could serve as a key mechanism for controlling this dual temporal process.(Johnson et al. 2025)

Complementing this rapid PTM-based regulation, we observed a slower regulatory method through substantial changes in the abundance of OpcA and G6PDH proteins when comparing diel cycle adapted cells at noon (cells rely on photosynthesis as the source of reducing power) to cells at subjective dusk (cells prepare for light being unavailable). Indeed, our previously published analysis of the *S. elongatus* transcriptome under diel conditions showed that both genes, *opcA* and *zwf,* encoding the G6PDH protein, were among the most differentially expressed genes between dusk (8:00 p.m.) and noon (12:00 p.m.) time points under identical light conditions.(Gilliam et al. 2025) Therefore this slower, abundance-based control mechanism has dramatic impact on cellular bioproduction function while operating on an hourly timescale. Because the PTM switching process is both reversible and localized, it enables a layered regulatory hierarchy: an ultrafast layer (ms–s), where cysteine redox toggles OpcA/G6PDH activity and, by extension, NADPH supply; and a slow layer (h–d) where changes in gene expression alter absolute enzyme abundance, locking in the metabolic state established by earlier redox cues. Moreover, it is possible that an intermediate layer (min–h) exists, where metabolite feedback (*e.g.*, accumulation of 6-phosphogluconate, ATP/ADP ratio) fine-tunes flux partitioning between the OPPP pathway and glycolysis/CBB.(Esposito 2016; Jiang et al. 2022; Preiser et al. 2019; Xu et al. 2024)

Such temporal partitioning ensures that cyanobacteria can respond to rapid environmental fluctuations through immediate NADPH mobilization while simultaneously adapting their proteome to sustained environmental changes. From a systems perspective, OpcA therefore acts as an integrator that coordinates reducing power homeostasis with carbon, nitrogen, and redox metabolism in real time, providing a paradigm for how single redox switches can orchestrate global pathway crosstalk with the primary goal of maintaining cellular reducing power balance.

*PTM-Psi integrated with redox proteomics provides a powerful approach for detecting cellular states by linking molecular details to systems modeling*

High-coverage redox proteomics now delivers quantitative site-occupancy data for thousands of cysteine residues, but converting these datasets into mechanistic insight remains challenging. PTM-Psi closes this gap by providing physics-based structural context for each modified site. The workflow is conceptually straightforward. Looking forward, integrating PTM-Psi/redox-proteomics data with machine-learning frameworks and time-resolved proteomics will enable real-time tracking of cellular states, guiding interventions ranging from developing stress-resilient and high productivity cyanobacterial strains with optimized reducing power management to therapeutic modulation of redox-sensitive enzymes in human disease.

## CONCLUSIONS

This study demonstrates that thiol PTMs function as allosteric switches that regulate cyanobacterial reducing power balance and carbon metabolism through structural modulation of the OpcA-G6PDH complex. Our integrative PTM-Psi/redox proteomics approach reveals mechanisms by which redox-sensitive cysteine modifications likely enable the rapid metabolic response of *S. elongatus* PCC 7942 to fluctuating environmental conditions.

At the molecular level, our simulations indicate that specific cysteine PTMs (C162-C174 and C380-C386 disulfide bonds, plus C398 glutathionylation) in OpcA protein function as conformational switches that appear to allosterically regulate G6PDH catalytic efficiency. These modifications promote optimal "open" gate conformations in three of four G6PDH subunits while potentially maintaining critical hydrogen bond networks (ASP33-ARG243, GLU241-ARG243) essential for active site integrity. The predicted 10 kcal/mol enhancement in OpcA-G6PDH binding energy upon PTM formation provides a computational basis for allosteric activation.

Together with a systems-level modeling, our data suggest hierarchical temporal regulation across different timescales: fast PTM switching potentially coupling redox state to NADPH production, and significantly slower OpcA and G6PDH proteins abundance-driven mechanisms. This multi-layered architecture may ensure both rapid environmental responsiveness and sustained metabolic optimization.(Johnson et al. 2025)

The strategic positioning of all cysteine residues within the G6PDH-binding domain and conservation of gate residues across cyanobacterial species (while C398 remains unique to *S. elongatus*) suggest evolutionary pressure for redox-sensitive metabolic control that may be essential for survival in fluctuating light environments.

Methodologically, this work demonstrates the power of integrating computational modeling with quantitative redox proteomics to investigate allosteric mechanisms. The successful prediction of PTM-induced conformational changes and their experimental validation through site-specific redox proteomics establishes this integrated approach as highly promising for investigating dynamic protein modifications.

The proposed gate mechanism could prove conserved across G6PDH enzymes, suggesting bioengineering applications for controlling NADPH production, rational manipulation of redox balance, and carbon metabolism in phototrophs for bioeconomy applications, with potential secondary applications in metabolic disease therapeutics. This study provides a framework for understanding how single amino acid modifications responsive to ROS may orchestrate global metabolic reprogramming through computational prediction of allosteric control in enzyme complexes.

## MATERIALS AND METHODS

*Global and redox proteome experiments*

Global and redox proteome experiments were conducted to investigate the redox state of cysteine residues in the OpcA protein of the bioproduction-relevant *S. elongatus* PCC 7942 cscB/SPS strain during light/dark transitions and diel circadian cycles. Data analysis was then performed against the *S. elongatus* PCC 7942 UniProt database.(Consortium 2024) Detailed procedures are outlined in Supporting Information.

*Computational study using PTM-Psi*

We used PTM-Psi to study the impact of PTM on the structure and function of OpcA.(Mejia-Rodriguez et al. 2023; Samantray et al. 2025) A workflow of PTM-Psi, a Python package to facilitate the computational investigation of post-translational modification on protein structures and their impacts on dynamics and functions, is provided in **Figure 6**. PTM-Psi is a user package that we previously developed to automate the various computational tasks, including building systems with multiple combinations of PTM on target proteins, implementing and validating force fields using quantum mechanics (QM) approaches for the non-standard PTM-ed amino acids, performing MD simulations, and analyzing simulation trajectories. Detailed procedures are outlined in Supporting Information as well as in our previous publication on PTM-Psi.(Mejia-Rodriguez et al. 2023; Samantray et al. 2025)

## SUPPLEMENTARY MATERIAL DESCRIPTION

Detailed procedures for experimental and computational approach, Amino acid sequence of OpcA and G6PD, Structural comparisons between their predicted and previous 3D structures, multiple sequence alignment, interaction energy between G6PDH and OpcA, RMSF of the OpcA-G6PDH complex, and native contacts from residues in OpcA to those for the gate of G6PDH.

## ACKNOWLEDGEMENTS

This work is mainly supported by the NW-BRaVE for Biopreparedness project funded by the U. S. Department of Energy (DOE), Office of Science, Biological and Environmental Research program, under FWP 81832. A portion of this research is supported by the Predictive Phenomics Initiative at Pacific Northwest National Laboratory. An award for computer time was provided by the ASCR Leadership Computing Challenge (ALCC) program. This research also used resources of the Oak Ridge Leadership Computing Facility, which is a DOE Office of Science User Facility supported under Contract DE-AC05-00OR22725. A part of this research was performed on a project award ("Enhancing biopreparedness through a model system to understand the molecular mechanisms that lead to pathogenesis and disease transmission") from the Environmental Molecular Sciences Laboratory (EMSL), a DOE Office of Science User Facility sponsored by the Biological and Environmental Research program. Specifically, part of the computational work in this paper was performed using the Molecular Sciences Computing Facility at EMSL. Pacific Northwest National Laboratory is a multi-program national laboratory operated by Battelle for the DOE under Contract No. DE-AC05-76RL01830.

**FIGURES**

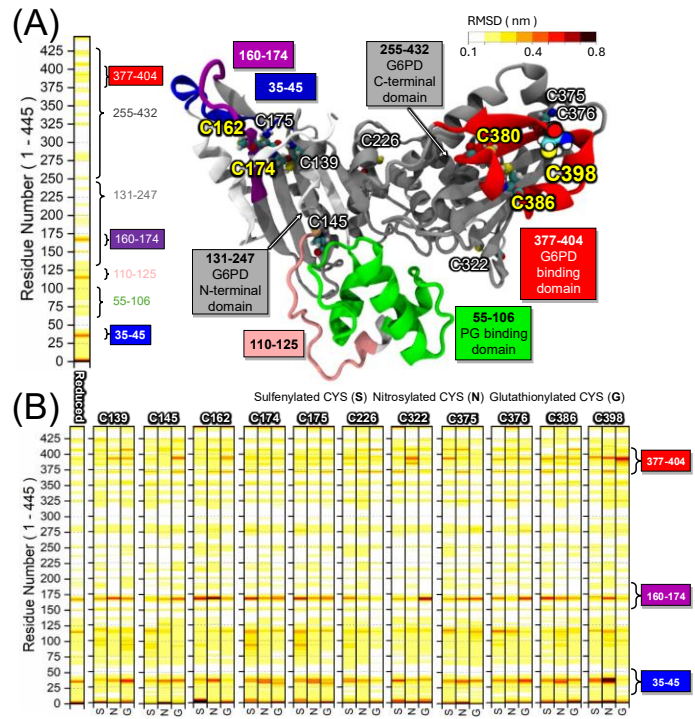

**Figure 1. (A)** RMSF analysis of the reduced OpcA showing regional flexibility patterns. A representative snapshot highlights putative functionally important regions: peptidoglycan binding site (green), G6PDH binding site (red), and terminal domains (gray), with flexible regions identified by MD simulations marked in purple, pink, and blue. **(B)** RMSF comparison between reduced and PTMed OpcA, where specific cysteine sites were replaced with sulfenylated (S), nitrosylated (N), or glutathionylated (G) cysteines. Darker red indicates higher RMSF and more flexible regions.

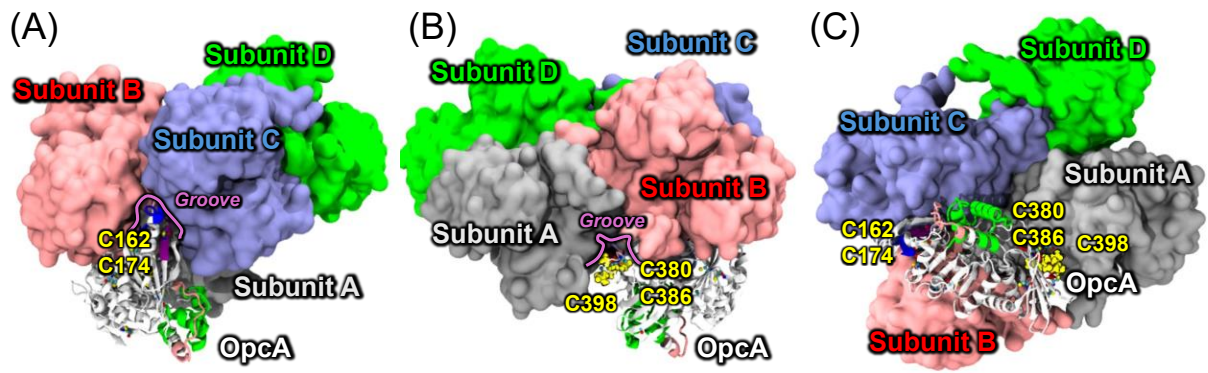

**Figure 2**. A predicted structure of the OpcA–G6PDH complex shown as a side view. **(A and B)** Side views highlighting close contacts between flexible regions I (residues 35-45, blue) and III (residues 160-174, purple), the putative G6PDH binding site (residues 377-404, red) in OpcA, and the G6PDH tetramer. Grooves between two G6PDH subunits that establish close contact with OpcA are highlighted. **(C)** Bottom view showing the OpcA monomer interacting with three G6PDH subunits (A – C).

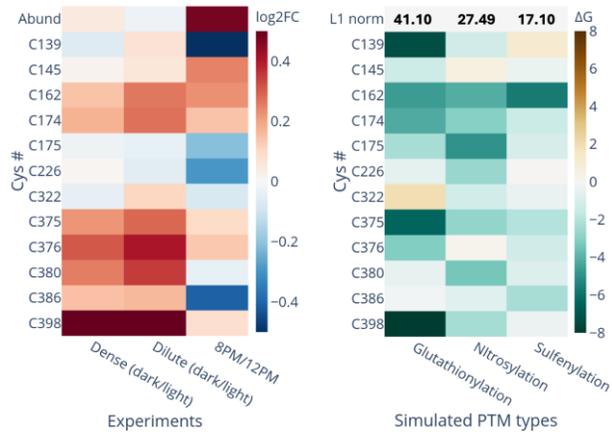

**Figure 3**. Experimental redox proteomics and computational free energy analysis of cysteine PTMs in OpcA. **(Left Panel)** Changes in oxidation levels of individual cysteine sites and OpcA protein abundance (labelled as "Abund") measured by redox proteomics. Heatmap shows fold changes in cysteine oxidation when light-adapted cyanobacterial cells are switched to dark for 2 h either under dense (high cell density with increased self-shading) or dilute (low cell density with reduced self-shading) culture conditions, and when comparing diel cycle-adapted cells at dusk (8:00 p.m.) versus noon (12:00 p.m.) under similar light conditions. Blue indicates a decrease; red indicates an increase in oxidation. **(Right Panel)** The RBFE for transformation from three different PTM-modified cysteine to reduced cysteine. More negative values indicate that a given PTM is energetically more favorable. The L1 norm is the calculated Manhattan distance of all the free energy changes from site modifications of corresponding PTM types. This L1 norm measures the overall effects of different PTM types on all captured sites.

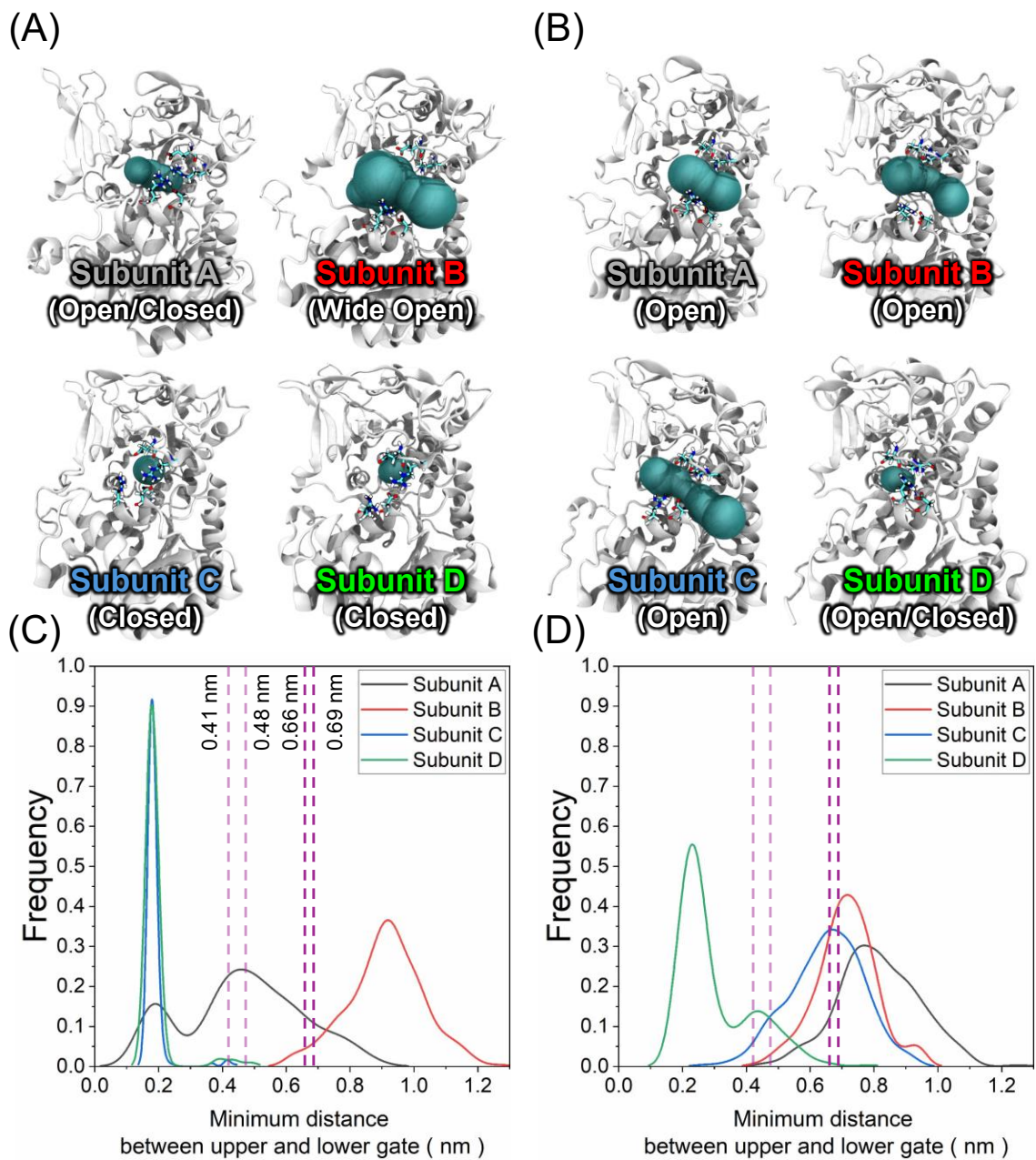

**Figure 4.** Representative snapshots of each subunit in the G6PDH complex in the **(A)** reduced and **(B)** oxidized cases with PTMs. The void area at the active site pocket is highlighted by cyan regions. Relative frequencies of the gate configurations in the **(C)** the reduced and **(D)** oxidized cases with PTMs. Distributions of subunits A, B, C, and D are highlighted in black, red, blue, and green, respectively. Purple vertical dashed lines represent the distances between the upper and lower parts of the gate for G6PDH protein complexes from different cyanobacteria.

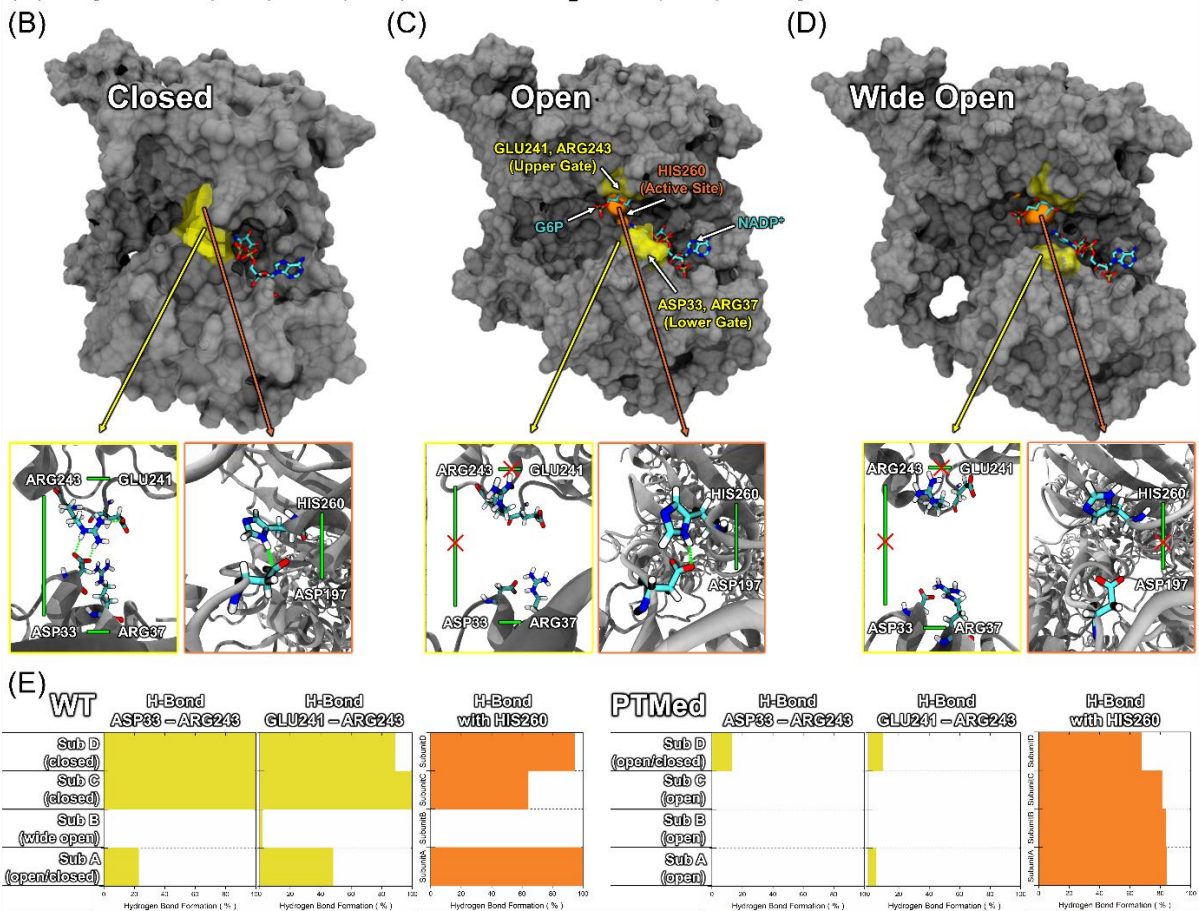

**Figure 5. (A)** The proposed reaction occurring at the active site of G6PDH. Representative snapshots of the G6PDH protein with three different gate conformations: **(B)** closed, **(C)** open, and **(D)** wide-open. The substrate and cofactor, G6P and NADP$^+$, were taken from the human G6PDH and docked into our system to better understand how reactants can bind to the active sites. **(E)** Percentage of hydrogen bond formation between amino acids in each G6PDH subunit involved in the gate configuration (yellow) and the stability of the active site (orange).

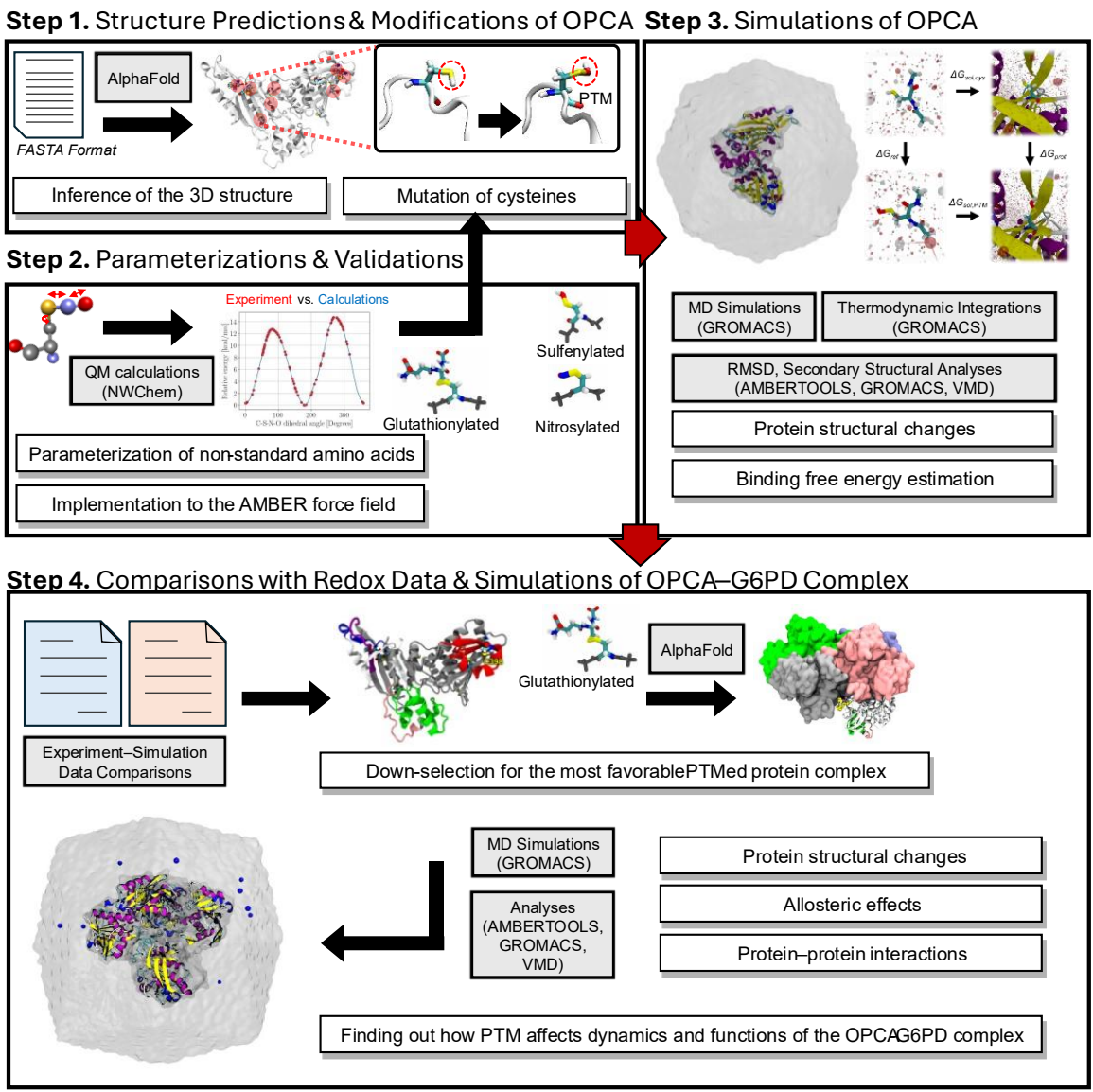

**Figure 6.** A workflow depicting the sequence of steps in this study. Steps 1 to 3 illustrate the automated steps of this study using the PTM-Psi package: from protein structure inference and parameterizations for the simulation, to actual MD simulations of the OpcA protein using PTM-Psi. Step 4 is a further step of this research, including comparisons between experimental and computational data, down-selection of representative cases for simulations of the OpcA-G6PDH complex, and relevant simulations and analyses.

# SUPPORTING INFORMATION

# Thiol post-translational modifications modulate allosteric regulation of the OpcA-G6PDH complex through conformational gate control

Hoshin Kim[1, #, *], Song Feng[2, #], Pavlo Bohutskyi[2,3], Xiaolu Li[2], Daniel Mejia-Rodriguez[1], Tong Zhang[2], Wei-Jun Qian[2], and Margaret S. Cheung[4,5*]

[1]Physical Sciences Division, Physical and Computational Sciences Directorate, Pacific Northwest National Laboratory, Richland, Washington, USA

[2]Biological Sciences Division, Earth and Biological Sciences Directorate, Pacific Northwest National Laboratory, Richland, Washington, USA

[3]Department of Biological Systems Engineering, Washington State University, Pullman, Washington, USA

[4]Environmental Molecular Sciences Laboratory, Richland, Washington, USA

[5]University of Washington, Seattle, Washington, USA

[#] shares co-first authorship of this article.

[*] shares co-corresponding authorship of this article: hoshin.kim@pnnl.gov, margaret.cheung@pnnl.gov

## Detailed procedures for experimental and computational approaches

*Global and redox proteome experiments*

Two cultures, one adapted to continuous light and another to a 14:10 h light-dark cycle, were grown under ~200 µmol m$^{-2}$ s$^{-1}$ photon flux density at 29±2°C with 2% $CO_2$ in modified BG-11 medium to maintain an $OD_{750}$ of ~0.2. Experiments included redox analysis after transitioning from light to dark at high (OD750 1.0) and low (OD750 0.08) cell densities, as well as cysteine oxidation profiling at noon and dusk in diel-adapted cultures. Protein samples were prepared by cold TCA precipitation, lysed with urea/EDTA/SDS/NEM, enriched using resin-assisted capture (RAC) for oxidized cysteine residues, and analyzed by LC-MS/MS using isobaric labeling for quantification. Global and redox proteome experiments were performed to investigate the redox state of cysteine residues in the OpcA protein using the *S. elongatus* PCC 7942 cscB/SPS strain. This strain was selected as a bioproduction-relevant model system, and the experiment was designed to mimic physiological transitions where OpcA is involved in modulating cellular reducing power generation and carbon flux by activating the OPPP. We established two distinct *S. elongatus* cultures: one adapted to continuous light and a second diel culture that was adapted to a 14:10 h light-dark cycle for approximately 6 months. Both cultures were cultivated under ~200 µmol m$^{-2}$ s$^{-1}$ photon flux density at 29±2 °C and supplied with 2% $CO_2$. The continuous light culture was grown in a modified BG-11 medium containing 0.09 g L$^{-1}$ Yeast Nitrogen Base without amino acids and ammonium sulfate (H26271.36, Thermo Fisher Scientific Inc., USA), 0.264 g L$^{-1}$ of $(NH_4)_2SO_4$ (J64180.A1, Thermo Fisher Scientific Inc., USA), and 0.174 g L$^{-1}$ of $K_2HPO_4$, (60,356, Sigma-Aldrich, USA), (I56000, RPI, USA). Cultures were maintained in the exponential growth phase through regular dilution with fresh medium to maintain an $OD_{750}$ of approximately 0.2.

To investigate how light-dark transitions and day-night dynamics affect the redox state of the OpcA protein, we performed three complementary experiments. The first two experiments, one with *S. elongatus* cultures at high cell density ($OD_{750}$ 1.0) and another at low cell density ($OD_{750}$ 0.08), measured changes in the redox state of OpcA after transitioning from light to dark conditions. The third experiment measured changes in the redox state of OpcA in diel-synchronized cultures at different points of the circadian cycle, specifically at dusk (8:00 p.m.) versus noon (12:00 p.m.) under identical light conditions.

For the light-to-dark transition experiments, cultures were cultivated under continuous light from $OD_{750}$ 0.15 to $OD_{750}$ 1.0 (late mid-exponential phase) over 4 days. After reaching the desired optical density, half of the culture was divided into individual culture flasks that served as experimental replicates for light-to-dark transition under high-density conditions. The remaining half was diluted from $OD_{750}$ of 1.0 to 0.08 by using filter-sterilized spent medium from the same cell culture. Both dense and dilute cultures were then exposed to the same light intensity for 0.5 h, and light condition samples were collected for both densities. Importantly, individual cells in the dilute culture (OD 0.08) received much higher light irradiance due to reduced self-shading compared to cells in the dense culture (OD 1.0). Subsequently, the cultures were transferred to darkness for 2 h, and dark condition samples were collected.

For the circadian cycle comparison experiment, we cultured two groups of diel cycle-adapted cells and harvested one group at 12:00 p.m. (noon) and another group at 8:00 p.m. (dusk). A detailed description of this experiment was reported earlier.(Gilliam et al. 2025) For each biological replicate, 100 mL of culture was divided into two 50 mL aliquots and centrifuged at 4,700 g for 5 min at 4 °C. The supernatant was decanted, and cell pellets were flash frozen in liquid nitrogen and stored at –80 °C. Dark treatment samples were maintained in darkness during all processing steps. Samples were then processed as described previously (Guo et al. 2014c). Briefly, cell pellets were treated with 10% cold trichloroacetic acid on ice, then incubated in 250 mM HEPES (pH 7.5) containing 8 M urea, 10 mM ethylenediaminetetraacetic acid (EDTA), 0.5% sodium dodecyl sulfate (SDS), and 100 mM N-ethylmaleimide (NEM) for 2 h, then lysed by bead beating. The extracted protein was precipitated and washed with cold acetone, then reduced with dithiothreitol. The reduced protein was further processed with a resin-assisted capture (RAC) protocol to selectively enrich proteins with oxidized cysteine residues.(Guo et al. 2014b) A detailed experimental procedure for the redox proteome was described earlier.(Gaffrey et al. 2021)

Simultaneously, an aliquot was taken from each protein sample before resin-assisted capture enrichment to quantify global protein expression levels. Isobaric labeling was used for quantification. The enriched samples and global protein samples were analyzed by a nanoAcquity UPLC system (Waters) coupled to a Q-Exactive HF-X Orbitrap mass spectrometer (Thermo Scientific, San Jose, CA) with a 120-minute liquid chromatography (LC) gradient. Full mass spectrometry (MS) scans were acquired in the range of m/z = 400 – 1800. MS/MS was performed in data-dependent mode. The parent ion tolerance was 20 ppm. LC-MS/MS raw data were searched against the *S. elongatus* PCC 7942 UniProt database (downloaded March 8, 2022) using MS-GF+.(Guo et al. 2014b; Kim and Pevzner 2014) A detailed experimental procedure for this process was described previously.(Guo et al. 2014b)

*MD simulations of OpcA with PTMs*

The PTM-Psi package provides a streamlined, end-to-end automated workflow from a protein sequence to trajectory analysis. Here we break down each step for clarity. First, we obtained a primary protein (FASTA) sequence of OpcA from *S. elongatus* strain PCC 7942 (hereafter, *S. elongatus*) from UniProt (ID: Q54709, OpcA_SYNE7) and used AlphaFold 2(Jumper et al. 2021) to infer its 3D structure (**Figure S1**). The predicted structure was compared with the experimentally verified OpcA from a different cyanobacterium, *Synechocystis sp.* PCC 6803,(Doello et al. 2024) using the MM-align tool to get the TM-score for assessing protein similarities.(Mukherjee and Zhang 2009) Two structures with TM-scores between 0.5 and 1.0 have considerably high structural similarities(Zhang and Skolnick 2004) and, as illustrated in **Figure S2**, the predicted OpcA structure showed good alignment with the cryo-EM-based structure.(Doello et al. 2024) Using PTM-Psi, all cysteine residues were considered PTM sites and modified to three different types of PTM: sulfenylated, nitrosylated, and glutathionylated. The force field parameters of these non-standard amino acids, including both bonded and non-bonded terms, were carefully parameterized and validated using the QM approach.(Mejia-Rodriguez et al. 2023) These parameters were integrated with the PTM-Psi package, extending the capabilities of the AMBER99SB force field.(Hornak et al. 2006) Based on the number of cysteine sites and three different PTM types, we built 34 different systems (11 cysteines × 3 PTM types + 1 completely reduced OpcA) and created three replicas per system for improved sampling from the MD simulations. A total of 102 unique cases were then solvated in a truncated dodecahedron periodic box filled with TIP3P water molecules(Jorgensen et al. 1983) with monovalent ions based on Joung and Cheatham ion parameters(Joung and Cheatham 2008) to neutralize the charge. GROMACS 2024.3(Abraham 2024) was used for the MD simulations. The equilibrations began with solvent minimization while keeping the protein atoms restrained at 1,000 kJ mol$^{-1}$·nm$^{-2}$. Additional multistep minimizations were subsequently performed, where harmonic restraints on the protein atoms were gradually reduced from 500 to 200, 100, 10, 5, and 1 kJ mol$^{-1}$·nm$^{-2}$. Then, the system was equilibrated at 300 K for 500 ps in the constant-temperature, constant-volume (NVT) ensemble using velocity rescaling with a stochastic term thermostat,(Bussi et al. 2007) followed by 500 ps equilibration at 1 bar and 300 K under a constant-temperature, constant-pressure (NPT) ensemble using the stochastic cell rescaling (C-rescale) method for pressure control.(Bernetti and Bussi 2020) The final minimization was carried out without harmonic restraints. Afterward, the system was re-equilibrated without any restraints using the same procedures for the NVT and NPT ensembles. The systems were then propagated for 100 ns at 300 K and 1 bar using the Parrinello–Rahman pressure coupling method(Parrinello and Rahman 1981) along with the particle mesh Ewald method for long-range electrostatic interactions,(Darden et al. 1993) and employing the linear constraints solver (LINCS)(Hess 2008; Hess et al. 1997) to constrain all bonds to hydrogen atoms. All simulations adopted a timestep of 2 fs.

Subsequently, we performed thermodynamic integrations (TI) of all distinct systems to compute the free energy cost associated with transforming a protonated cysteine (CYS) state to its sulfenylated, nitrosylated, or glutathionylated form. In the TI approach, non-bonded and bonded interactions were changed via a two-step method where electrostatic contributions were decoupled first, followed by van der Waals, masses, and bonded terms. These terms were gradually decoupled by increasing the coupling parameter ($\lambda$) from 0.00 to 1.00 across 13 discrete windows ($\lambda$ = 0.00, 0.05, 0.10, 0.20, …, 0.90, 0.95, 1.00), for each decoupling step. Thus, a total of twenty-six 1 ns trajectories were employed for the calculation. TIs were also performed for the same transitions in TIP3P water,(Jorgensen et al. 1983) which were used as reference values to obtain the relative binding free energies (RBFE). Free energy differences were obtained using Bennett's acceptance ratio method(Bennett 1976) as implemented in the g_bar module of the GROMACS simulation package.(Abraham 2024) The RBFEs were calculated as the differences between the free energy costs observed in the protein systems and those in water. This alchemical process was incorporated in PTM-Psi, with further details provided in our previous papers.(Mejia-Rodriguez et al. 2023; Samantray et al. 2025)

*MD simulations of the Reduced and PTM-ed OpcA-G6PDH binary complexes*

For simulating the OpcA-G6PDH binary complex, a protein sequence of G6PDH from the same strain, *S. elongatus,* was first obtained from the UniProt database, P29686, G6PDH_SYNE7 (**Figure S1**). We used AlphaFold 3(Abramson et al. 2024) to predict the hetero-oligomeric structure composed of a monomer of OpcA and a tetramer of G6PDH proteins. MM-align(Mukherjee and Zhang 2009) was used to assess the structural similarity between predicted and cryo-EM-based complex structures, demonstrating good structural similarity (**Figure S2**). Based on our computational RBFE results, along with a previous study(Doello et al. 2024) and experimental redox proteome data, we downselected cysteine sites that tend to be PTMed under dark conditions and created two distinct complexes for further simulations: the one is a completely reduced complex where no disulfide bonds or thiol PTMs are formed, the other is a PTMed complex with two disulfide bonds, C162-C174 and C380-C386, as well as glutathionylated C398. The aforementioned procedures were applied for the MD simulations of these two OpcA-G6PDH binary complexes.

*Computational Analyses*

Post-processing of trajectories, including translation of the protein to the center of the periodic box, was done using CPPTRAJ in AMBERTOOLS(Roe and Cheatham 2013; 2018) and GROMACS analysis tools.(Abraham 2024) Root mean square fluctuations (RMSF), radius of gyration ($R_g$), and hydrogen bonds, native contacts for minimum distance calculations, as well as long-range contacts, were obtained using the GROMACS rmsf and gyrate module,(Abraham 2024) and the hbond and nativecontacts commands in the CPPTRAJ module, respectively.(Roe and Cheatham 2013; 2018) The STRIDE algorithm was used for predicting secondary structure contents.(Heinig and Frishman 2004) Prediction of catalytic pockets was performed using Caver 3.02.(Chovancova et al. 2012; Pavelka et al. 2016) gmx_MMPBSA(Valdés-Tresanco et al. 2021) was employed to assess the interactions between OpcA and the G6PDH complex. We used Clustal Omega(Sievers et al. 2011) for multiple sequence alignment. Simple substrate docking was done via Discovery Studio Visualizer (DSV).(BIOVIA 2019) All visualizations, including the rendered image of OpcA, G6PDH, and the complex, were obtained via VMD 1.9.4(Humphrey et al. 1996) and DSV.(BIOVIA 2019) Protter was used to visualize the protein sequence of OpcA and G6PDH proteins.(Omasits et al. 2013)

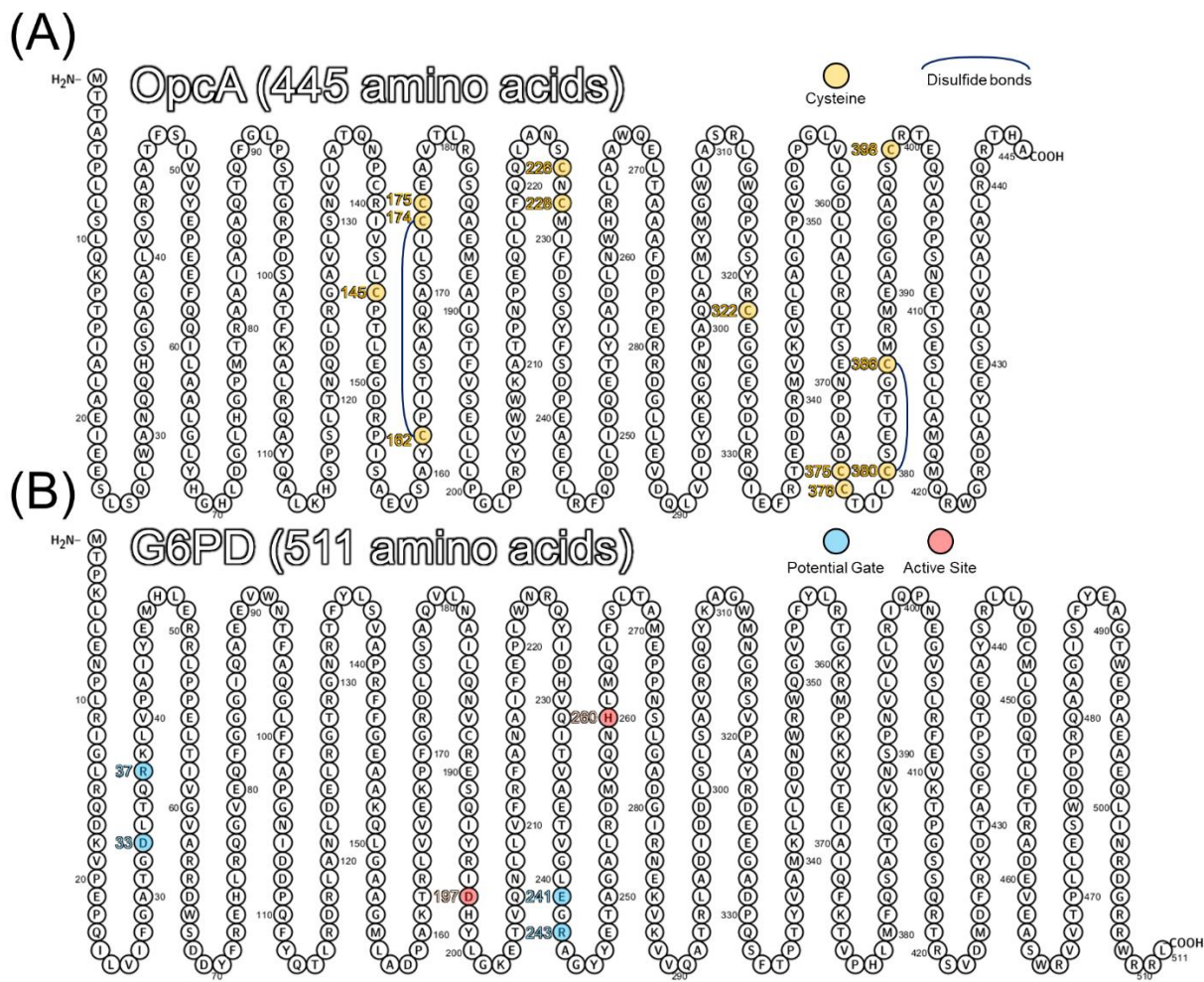

**Figure S1.** (A) Amino acid sequence of the OpcA monomer with the locations of cysteines (highlighted in yellow circles) chosen to be PTMed. Cysteine residue pairs that form disulfide bonds are highlighted by blue curved lines. (B) Amino acid sequence of the G6PDH monomer with the locations of possible residues involved in the gate configurations (highlighted in blue circles) and the active site (highlighted in red circles).

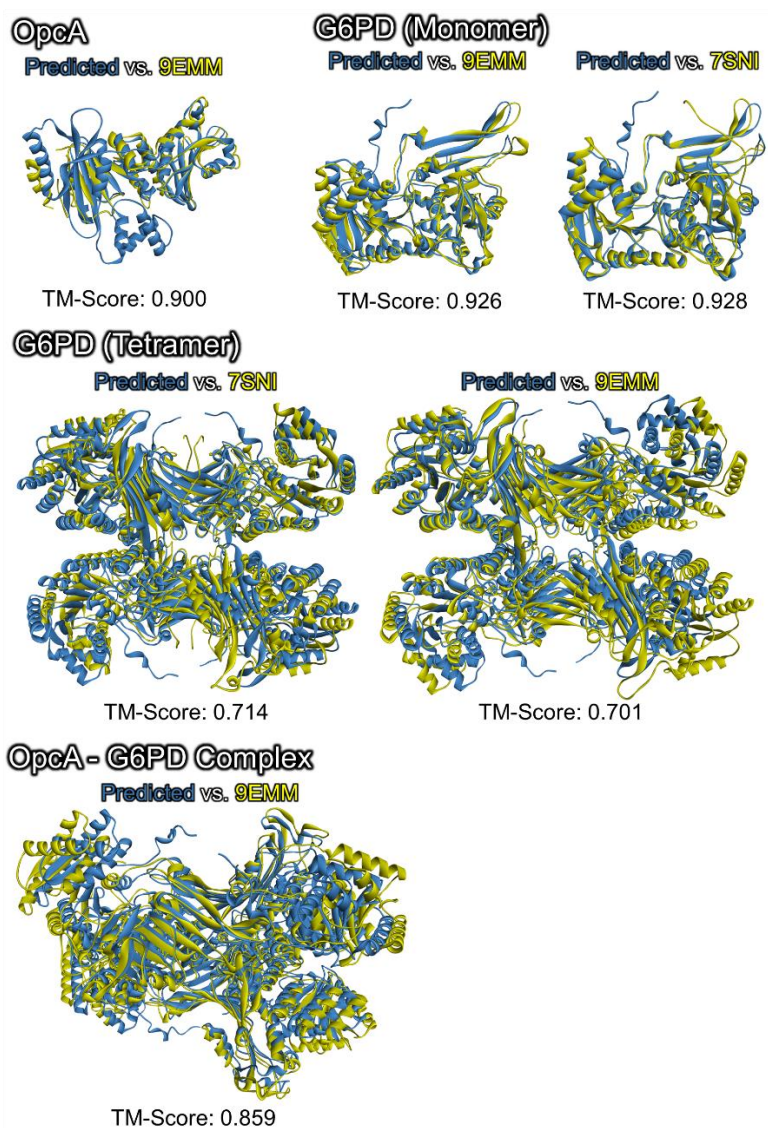

**Figure S2.** Structural comparisons between the predicted structures of OpcA, G6PDH monomer, G6PDH tetramer, and the OpcA-G6PDH complex with their corresponding 3D structures available in the PDB database. The PDB ID: 9EMM contains the OpcA-G6PDH complex from the cyanobacterial strain *Synechocystis* sp. PCC 6803, while 7SNI represents only the G6PDH tetrameric complex from humans. The figure includes TM-scores for each comparison, ranging from 0 to 1, with 0 indicating a poor match and 1 indicating a perfect match.

## (A)

Cysteine residue # of OpcA from S. elongatus (SYNE7)

☐ Cysteine  ☐ Not conserved residue  ⎴ Disulfide Bond observed in both OpcA

```
                                                        145
OPCA_SYNE7
ATFKALRQAYQALKHS----------PSLTNQDLRGAVLSNVIATQNPCRIVSLCPTLEG     150
OPCA_SYNY3
EFKQALQTAFETAHREGNLLSTAERITKPYSPDLEGSGIADTIAASNPCRIITLCPTAED    180

OPCA_SYNE7     162      174 175
DRPISAEVSAYCPITSAKQASLICCEAVTLRGSQAEMEAIGTFVSELLLPGLPRYVWWKA    210
OPCA_SYNY3
DQGVQAQLSAYCPIQKTHQNTLICCEYITLRGTSDALERIGGVITELMLPTLPKYVWWKA    240

OPCA_SYNE7          224 226
TPNPEQLLFQQLANSCNCMIFDSSYFSDPEAEFLRFQDLIDQETYIADLNWHRLAAWQEL    270
OPCA_SYNY3
SPEAEYGLFQRLLSHADMIIVDSSIFNNPEQDLLQLAQLVNKPEAIADLNWSRLAPWQEL    300

OPCA_SYNE7                               322
TAAAFDPPERRDGLLEVDQLVIDYEKGNPAQALMYMGWIASRLGWQPVSYRCEGGEYDLR    330
OPCA_SYNY3
TAEAFDPPERRSAVGEIDQISIDYEKGNHAQALMYLGWVASRLQWTPVSYSYQPGVYEIH    360

OPCA_SYNE7                      375 376 380    386
QIEFRTEDDRMVKVELAGIPVGDPGLVLGDLIALRLTSENPDADCCTILCSETTGCMRME    390
OPCA_SYNY3
KIQFCAPNQRPIEAELAGLPLADTGQVLGDLISLKLGSTNTQAQCGTVLCSGTVGCMRME    420

OPCA_SYNE7  398
AGGGAQSCRTEQVAPPSNETSESLLAMQMQRWGRDALYEESLAVIAVALRQRTHA    445
OPCA_SYNY3
AGGGAQNYRVQQVTALDDQNTEQLLGRQLQRWGRDALYDESMAIVLAILQLSQAG    475
```

## (B)

Key Residue # for the Gate of OpcA from S. elongatus (SYNE7)
Active Site # of OpcA from S. elongatus (SYNE7)

☐ Conserved gate residue  ☐ Not conserved but positively charged residue  ☐ Conserved active site

```
G6PD_HUMAN
MAEQVALSRTQVCGILREELFQGD-AFHQSDTHIFIIMGASGDLAKKIYPTIWWLFRDG    59
G6PD_SYNE7                                  33   37
----------MTPKLLENPLRIGLRQDKVPEPQILVIFGATGDLTQRKLVPAIYEMHLER    50
G6PD_SYNY3
----------MVTLLENPFRTGLRQERTPEPLILTIFGASGDLTQRKLVPAIYQMKRER    49

G6PD_HUMAN
FGRDLQSSDRLSNHISSLFREDQIYRIDHYLGKEMVQNLMVLRFANRIFGPIWNRDNIAC    232
G6PD_SYNE7                  198
FGRDLSSAQVLNAILQNVCRESQIYRIDHYLGKETVQNLLVFRFANAIFEPLWNRQYIDH    229
G6PD_SYNY3
FGRDLSSAQSLNRVVQSVCKENQVYRIDHYLGKETVQNLMVFRFANAIFEPLWNRQFVDH    228

G6PD_HUMAN
VILTFKEPFGTEGRGGYFDEFGIIRDVMQNHLLQMLCLVAMEKPASTNSDDVRDEKVKVL    292
G6PD_SYNE7   241 243              260
VQITVAETVGLEGRAGYYETAGALRDMVQNHLMQLFSLTAMEPPNSLGADGIRNEKVKVV    289
G6PD_SYNY3
VQITVAETVGVEERAGYYESAGALRDMVQNHLMQLFCLTAMDPPNAIDADSIRNEKVKVL    288
```

**Figure S3.** Multiple sequence alignment of (A) OpcA (OPCA_SYNE7) and (B) G6PDH protein (G6PD_SYNE7) used in this study. For (A), only OpcA from the cyanobacterial strain *Synechocystis* sp. PCC 6803 (OPCA_SYNY3), where its reaction mechanisms were previously explored, is compared. Yellow highlights represent cysteine residues of interest, and red highlights represent the non-conserved region in the different cyanobacteria, highlighting that C398 is not conserved. Cyan and magenta in panel (B) highlight conserved residues at the upper and lower parts of the gate and active site residue, respectively. A blue-highlighted residue represents a non-conserved residue with a similar positively charged side chain compared to the ARG (LYS).

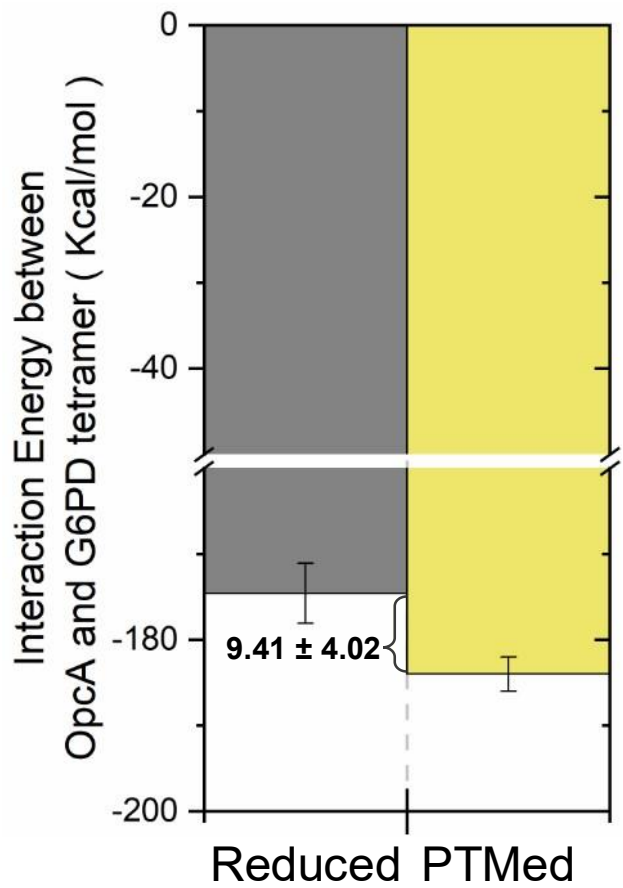

**Figure S4.** Interaction energy between the entire G6PDH subunits and reduced (gray) or PTMed (yellow) OpcA.

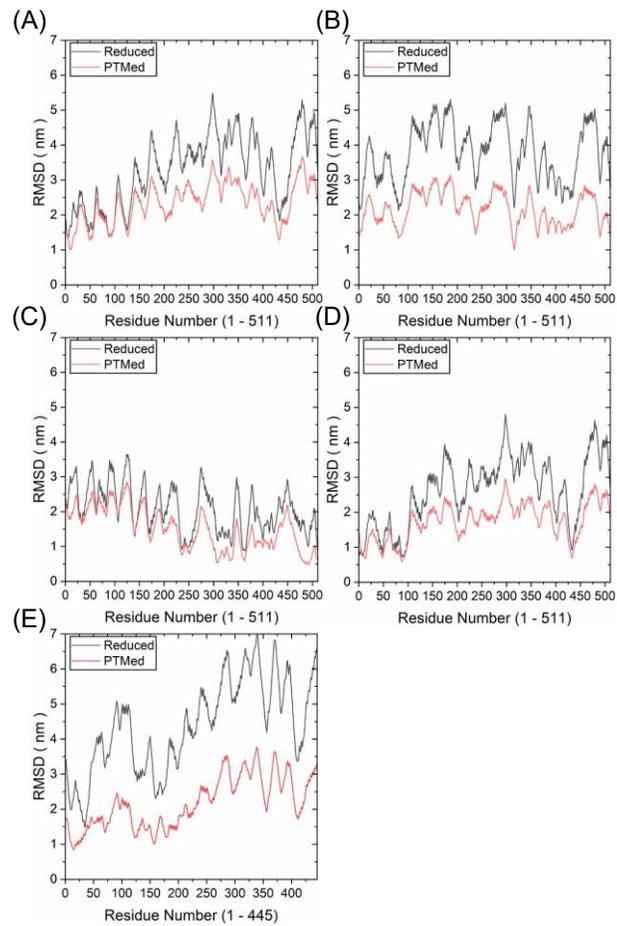

**Figure S5.** Residue-based root mean square fluctuation (RMSF) of (A-D) subunit A-D of G6PDH and (E) OpcA protein in the complex. Black and red represent the RMSF of the reduced and PTMed complex, respectively.

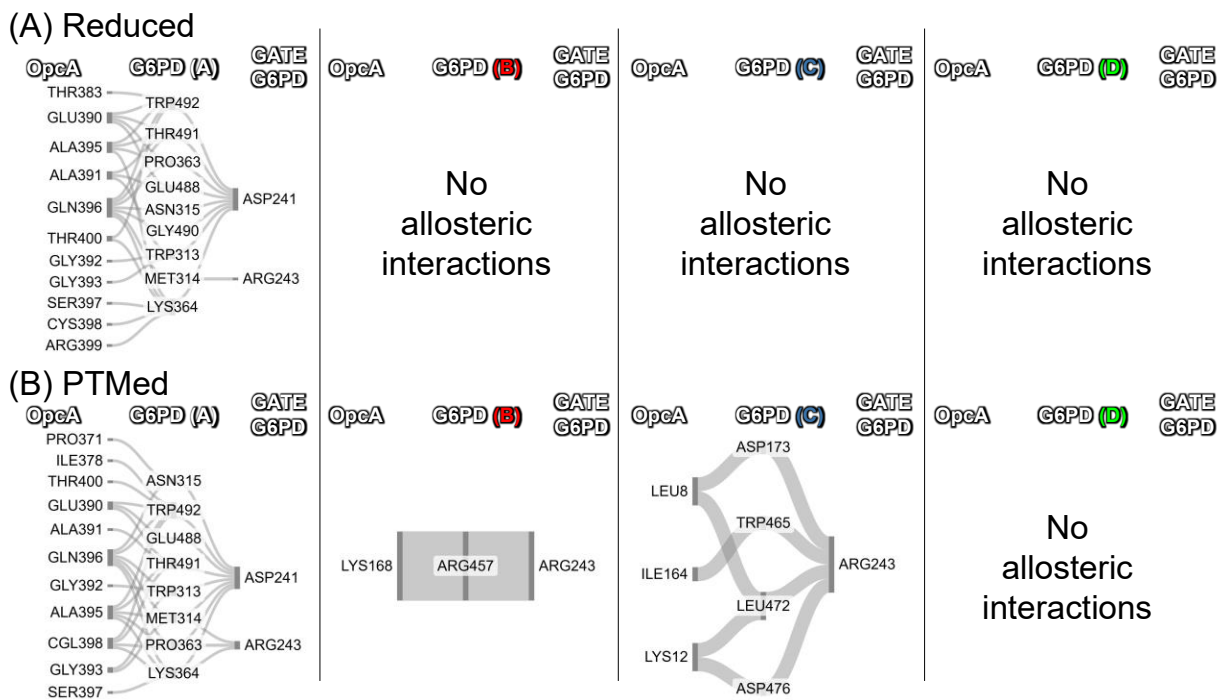

**Figure S6.** Native contacts from residues in OpcA, G6PDH, and the residues for the gate of G6PDH protein in (A) the complex with the reduced cysteines and (B) one with PTMed one. This flow represents how residues in OpcA can interact with residues at the gate of G6PDH and impact them allosterically.